 \documentclass[12pt,preprint]{aastex}

 \usepackage{emulateapj5}

\slugcomment{to appear in the Astrophysical Journal, Letters, 
2004 September 1}

\shorttitle{Premaximum Halt of Classical Novae}
\shortauthors{Hachisu \& Kato}

\begin{document}

\title{The nature of premaximum halts of classical nova
outbursts: \\ V723 Cassiopeiae and V463 Scuti}


\author{Izumi Hachisu}
\affil{Department of Earth Science and Astronomy, 
College of Arts and Sciences, University of Tokyo,
Komaba, Meguro-ku, Tokyo 153-8902, Japan} 
\email{hachisu@chianti.c.u-tokyo.ac.jp}

\and

\author{Mariko Kato}
\affil{Department of Astronomy, Keio University, 
Hiyoshi, Kouhoku-ku, Yokohama 223-8521, Japan} 
\email{mariko@educ.cc.keio.ac.jp}

%
%



\begin{abstract}
     We present a new interpretation of long premaximum halts of
nova outbursts.  For V723 Cas (Nova Cas 1995) and V463 Sct 
(Nova Sct 2000), we have reproduced light curves, excluding the
brightness maxima, starting from 
the long premaximum halt through the late decay phase of the outbursts
using a steady-state optically thick wind model.
When the hydrogen-rich envelope of the white dwarf (WD) is 
massive enough, the star expands to $\sim 100~R_\sun$ or over
and its surface temperature decreases to below 7000~K. 
At this supergiant mimicry stage, 
the changes in both the photospheric radius and the 
temperature are small against the large increase 
in the envelope mass.  These changes cause a saturation in visual 
magnitude that lasts a long time before it begins to decline. 
This saturation is known as the premaximum halt of 
a classical nova outburst.
The visual magnitude during the saturation is close to
the bolometric magnitude, which is an upper limit for a
given WD mass.  
Since the WD masses are estimated to be $0.59 ~M_\sun$ for V723 Cas
and $1.1 ~M_\sun$ for V463 Sct by fitting the decline rate of nova
light curves, we can determine the absolute magnitude of premaximum
halts.  It is a refined Eddington luminosity.  Thus, the premaximum
halt of a nova works as a standard candle.
\end{abstract}




\section{INTRODUCTION}
     Some classical novae have a flat part or stagnation 
of the optical light curve before it reaches the maximum.
Such a part of the optical light curve is called ``premaximum halt'' 
(e.g.,  Fig. 1.8 of Payne-Gaposchkin 1957; 
see also Fig. 5.1 of Warner 1995). 
The physical origin of the premaximum halt has not been fully 
understood yet, although a few ideas have been proposed.
The duration of a premaximum halt is closely related to the
nova speed class: the typical duration is a few hours in fast
novae \citep[e.g.,][]{war89} while it is up to several months
in some slow novae (HR~Del, V1548~Aql, V723~Cas, and DO~Aql).

     \citet{fri92} studied the past spectroscopic observations
of HR~Del during its long premaximum stage.  He showed the 
presence of an almost stationary photosphere with very low
velocities, unlike for the majority of classical novae.
He then suggested that the conditions of thermonuclear
runaway were just marginally satisfied (marginally unstable)
in HR~Del.  This interpretation is compatible with the estimated
white dwarf (WD) mass of $0.52~M_\sun$ \citep{bru82} or
$0.595~M_\sun$ \citep{kur88} in HR~Del because the development
of nova outbursts is very slow in these relatively
low mass WDs \citep[e.g.,][]{pri95}.
On the other hand, \citet{ori93} proposed another idea
of the premaximum halt, that is, a local (or partial) 
thermonuclear runaway on a white dwarf surface.

     Recently, \citet{kat02t} have revealed that the fast nova
V463~Sct (Nova Sct 2000) has a long, at least, 24 day premaximum
halt.  The decline rate of the visual magnitude is estimated to
be $t_2 = 15 \pm 3$ day, indicating a WD much more massive than
$0.5 - 0.6 ~M_\sun$.  This seems not to be consistent with
Friedjung's (1992) interpretation.  This long premaximum halt
also seems to go against the idea given by \citet{ori93} because
the propagation timescale of burning fronts is a few to several
days on WDs.

     Kato et al.'s (2002) result stimulated us to interpret
the premaximum halt purely from the theoretical side.
It is well known that the visual magnitude,
$M_V$, attains its peak value when a nova photosphere expands
greatly with the total luminosity being kept at the Eddington
luminosity, because the photospheric temperature decreases to
below $6000$~K while passing the visual band pass.
This makes a wide flat peak in the optical luminosity
corresponding to the premaximum halt, although 
it occurs only when the envelope mass is very massive.
In this Letter, based on an optically thick wind model of nova 
outbursts, we try to model the premaximum halt
for two recent objects, V723 Cas (Nova Cas 1995) and 
V463 Sct (Nova Sct 2000), because V723 Cas is the slowest nova
and V463 Sct is the fastest nova with a long premaximum halt. 
In \S 2, our optically thick wind model
is briefly introduced and modeling the premaximum halt is given
for V723 Cas. In \S 3 modeling the premaximum halt of V463 Sct is
also given.  Discussion and conclusions follow in \S 4.

\placefigure{hr_diagram_v723cas1995}

\section{Modeling of V723 Cas Premaximum Halt}

\subsection{Optically thick wind model}
     Photospheres of novae expand greatly
up to $\sim 100 ~R_\sun$ or larger.  In such a supergiant mimicry 
configuration of the WD envelope, the structure of the envelope
can be approximated by a steady-state wind \citep[e.g.,][]{kat94h}.  
Here we assume spherical symmetry.
Using the same method and numerical techniques as in \citet{kat94h},
we have obtained a sequence of steady-state wind solutions that
mimics a time evolution of the decay phase of nova outbursts. 
This sequence is characterized by a decreasing envelope mass,
because the hydrogen-rich envelope mass decreases in time
as a result of wind mass loss and nuclear burning.

     In order to calculate steady state wind solutions,
we solve a set of equations, i.e., the continuity, equation
of motion, radiative diffusion, and conservation of energy, 
from the bottom of the hydrogen-rich envelope through the photosphere,
under the condition that the solution goes through a critical point
of the steady state wind.  OPAL opacity is adopted.  Details of 
the computations have been published in \citet{kat94h}.  
The physical properties of these wind solutions have also been
published in many papers \citep[e.g.,][]{hac01kb, hkkm00, hkn96,
hkn99, hknu99, kat83, kat97, kat99}.
It should be noticed that a large number of meshes, i.e., 
more than 10,000 grids, are adopted for the wind solutions
with an expanded photosphere of $\sim 100 ~R_\sun$. 

     Figure \ref{hr_diagram_v723cas1995} shows such 
a sequence for a $0.59 ~M_\sun$ WD
with hydrogen-rich envelopes of the solar composition.
When the envelope expands greatly after the onset of 
unstable hydrogen burning, the nova reaches somewhere
on the sequence in the HR diagram, which depends on the envelope
mass at the ignition.  Here, two very early stages are marked
by A and B.  The envelope mass at epoch A is 
$\Delta M_A = 6.0 \times 10^{-5} M_\sun$ and that of epoch B is
$\Delta M_B = 5.1 \times 10^{-5} M_\sun$.  The wind mass
loss rates are $\dot M_{\rm wind, A}= 2.1 \times 10^{-5} M_\sun$~yr$^{-1}$
and $\dot M_{\rm wind, B}= 1.3 \times 10^{-5} M_\sun$~yr$^{-1}$,
respectively. 

     In a large part of the decay phase of novae the luminosity
is fairly constant as shown in Figure
\ref{hr_diagram_v723cas1995} even in the wind phase.
However, the visual luminosity attains its maximum value 
at the photospheric temperature of $\sim 6000$~K.  Such a 
low surface temperature is reached only when the envelope mass
is as massive as $\sim 6.0 \times 10^{-5} M_\sun$ for the 
$0.59~M_\sun$ WD.
The visual magnitude hardly varies between epoch A and epoch B  
so that the visual light curve becomes flat during the time when
the nova moves from epoch A to epoch B.  We have easily estimated 
the duration of such a flat part of the visual light curve by 
\begin{equation}
\Delta t = \int_A^B {{d M} 
\over {- (\dot M_{\rm wind}+ \dot M_{\rm nuc})}}
 \sim {{\Delta M_{\rm A} - \Delta M_{\rm B}} 
\over {\dot M_{\rm wind}}} \sim 200 {\rm ~day},
\label{premaximum_halt_duration}
\end{equation}
because the envelope mass is decreasing mainly 
by the wind mass-loss ($\dot M_{\rm nuc} \ll \dot M_{\rm wind}$),
where $\dot M_{\rm nuc}$ is the nuclear (hydrogen) burning rate and
$\dot M_{\rm wind}$ is the wind mass loss rate and we 
use $\dot M_{\rm wind}= 
(\dot M_{\rm wind, A} \dot M_{\rm wind, B})^{1/2}$. 
If the ignition mass is less massive than $\Delta M_B$, the nova
does not have a flat peak.

\placefigure{vmag_mmix_v723cas1995_early}

\placetable{standard_candle}

\subsection{Premaximum Halt of V723 Cas} 

     Figure \ref{vmag_mmix_v723cas1995_early} shows the light 
curve fitted with the V723 Cas observation.
Assuming the solar abundance of the envelope composition,
we have calculated 12 cases of the WD mass, i.e., 
0.56, 0.58, 0.59, 0.60, 0.65, 0.70, 0.80, 0.90,
1.0, 1.1, 1.2, and $1.3 ~M_\sun$ and found that
the $0.59 ~M_\sun$ WD model is the best-fit one for V723 Cas.

     The main difference between our model and the observation is
an oscillatory behavior that shows a strong (maximum) peak followed
by a few other sharp peaks, which are a bit darker than the maximum one.  
We interpret these peaks as pulsations of the nova
envelope as discussed by \citet{sch99}.  It should be noted here
that our steady state wind model cannot describe pulsations of 
the nova envelope because pulsations are not in a steady state.
Our sequence in Figures \ref{hr_diagram_v723cas1995} and 
\ref{vmag_mmix_v723cas1995_early} based on the steady state model
corresponds to the zeroth-order solutions when the pulsational
instability is suppressed.  

     Comparing our theoretical magnitude with the observation,
we have estimated the apparent distance modulus as
$(m-M)_V = 13.9$ for V723 Cas.  The interstellar
extinction toward V723 Cas have been estimated by many authors
but the values are scattered; that is, 
$E(B-V) = 0.20-0.25$ \citep{rud02},  
$0.29$ calculated from $A_V = 0.89$ \citep*{iij98},  
$0.45$ \citep{mun96},  
$0.57$ \citep{cho97},  
$0.60$ \citep*{gon96, ohs96},  
and $0.78$ \citep{eva03}, in increasing order.
The distance to V723 Cas is $d = 2.5$~kpc if we adopt $E(B-V)= 0.60$
while it is $d = 4.0$~kpc for $E(B-V)= 0.30$. 

\placefigure{vmag_mmix_v723cas1995}
\placefigure{rmag_v723cas_late_orbital}

\subsection{Light curve fitting in late phase}

     Figure \ref{vmag_mmix_v723cas1995} shows the entire phase
of the light curve of V723 Cas.  Our $0.59 ~M_\sun$ WD
model consistently follows the observation until 
JD 2,451,500 but gradually deviates more than a magnitude
after that.  In order to fit the light curve, we must 
consider the irradiation effects of the accretion disk 
and the companion star, supported by the orbital modulation 
that has gradually grown up to a magnitude in the visual band 
\citep[e.g.,][]{gor00, cho03}.  Including the irradiation
effects, we have calculated the orbital light curves.  
Here, we adopt a binary model consisting of a lobe-filling
companion of $0.5 ~M_\sun$ with the original surface temperature
of 4000~K, a $0.59 ~M_\sun$ WD, and an accretion disk around the WD.
The important parameters that determine the shape 
of the orbital light curve are
the size ($\alpha = R_{\rm disk}/ R_{\rm RL}$)
and thickness ($\beta = h_{\rm disk}/R_{\rm disk}$)
of the accretion disk and the inclination angle ($i$) of the orbit,
where $R_{\rm RL}$ is the effective radius of the Roche lobe,
$R_{\rm disk}$ is the disk radius, and $h_{\rm disk}$ is the
height of the disk 
\citep[see, e.g.,][for more detail]{hkkm00, hac01kb, hac04k}.
Here, we adopt $\alpha = 0.8$, 
$\beta = 0.04$ , and $i = 50\arcdeg$.
We adopt the ephemeris of
\begin{equation}
t(\mbox{HJD})= 2,450,421.4801 + 0.69325~E,
\label{new_ephemeris}
\end{equation}
at brightness minima \citep{gor00}. 
As shown in Figure \ref{rmag_v723cas_late_orbital},
the orbital light curve is successfully reproduced.
With this binary model, we are able to roughly follow
the entire phase of the light curve 
as shown in Figure \ref{vmag_mmix_v723cas1995}.

\placefigure{vmag_outburst_v463sct}

\section{Premaximum halt of V463 Sct}

     V463 Sct (Nova Sct 2000) is a fast nova with 
$t_2 = 15 \pm 3$ day \citep{kat02t}.  We adopt a carbon-oxygen
rich envelope of
$X= 0.35$, $Y= 0.33$, $C+O= 0.30$, and $Z=0.02$. 
Among seven sequences calculated for 
0.6, 0.7, 0.8, 0.9, 1.0, 1.1, and $1.2 ~M_\sun$ WDs,
three are plotted in Figure \ref{vmag_outburst_v463sct}.
The best fit one is that for the $1.1 ~M_\sun$ WD.
The envelope mass at the start
of the premaximum halt is $\Delta M = 6.8 \times 10^{-5} M_\sun$
for our $1.1 ~M_\sun$ WD model.
The apparent distance modulus is estimated to be
$(m-M)_V = 18.0$, so that the distance to V463 Sct is
$d = 12.7$~kpc if we adopt the interstellar extinction of 
$E(B-V) = 0.8$ \citep{kat02t}.

\placetable{properties_two_novae}

\section{Discussion}
\begin{figure}
\plotone{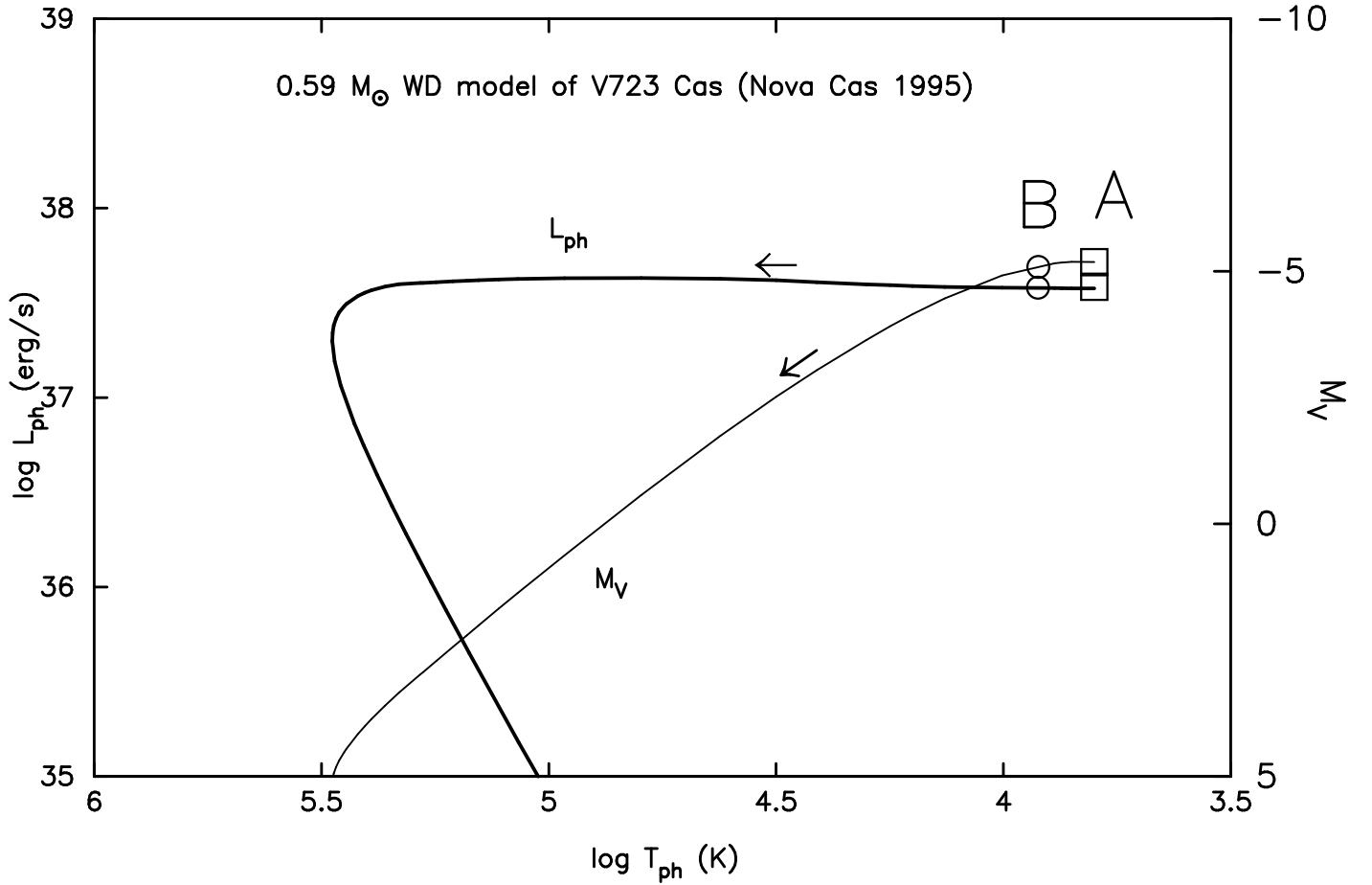}
\caption{
Our nova sequence of a $0.59 ~M_\sun$ WD plotted in the H-R diagram
together with its absolute visual magnitude.  
The nova moves along the arrow.
Two specific states are marked by A ({\it open square}) and 
B ({\it open circle}).  The envelope masses at epoch A and epoch B 
are $6.0 \times 10^{-5}$ and $5.1 \times 10^{-5} M_\sun$,
respectively.  The wind mass-loss rates are $2.1 \times 10^{-5}$ and
$1.3 \times 10^{-5} M_\sun$~yr$^{-1}$, respectively.
The absolute visual magnitude attains its maximum around
epoch A because the surface temperature decreases below $6000$~K
on the right hand side of epoch A.
\label{hr_diagram_v723cas1995}}
\end{figure}

\begin{figure}
\plotone{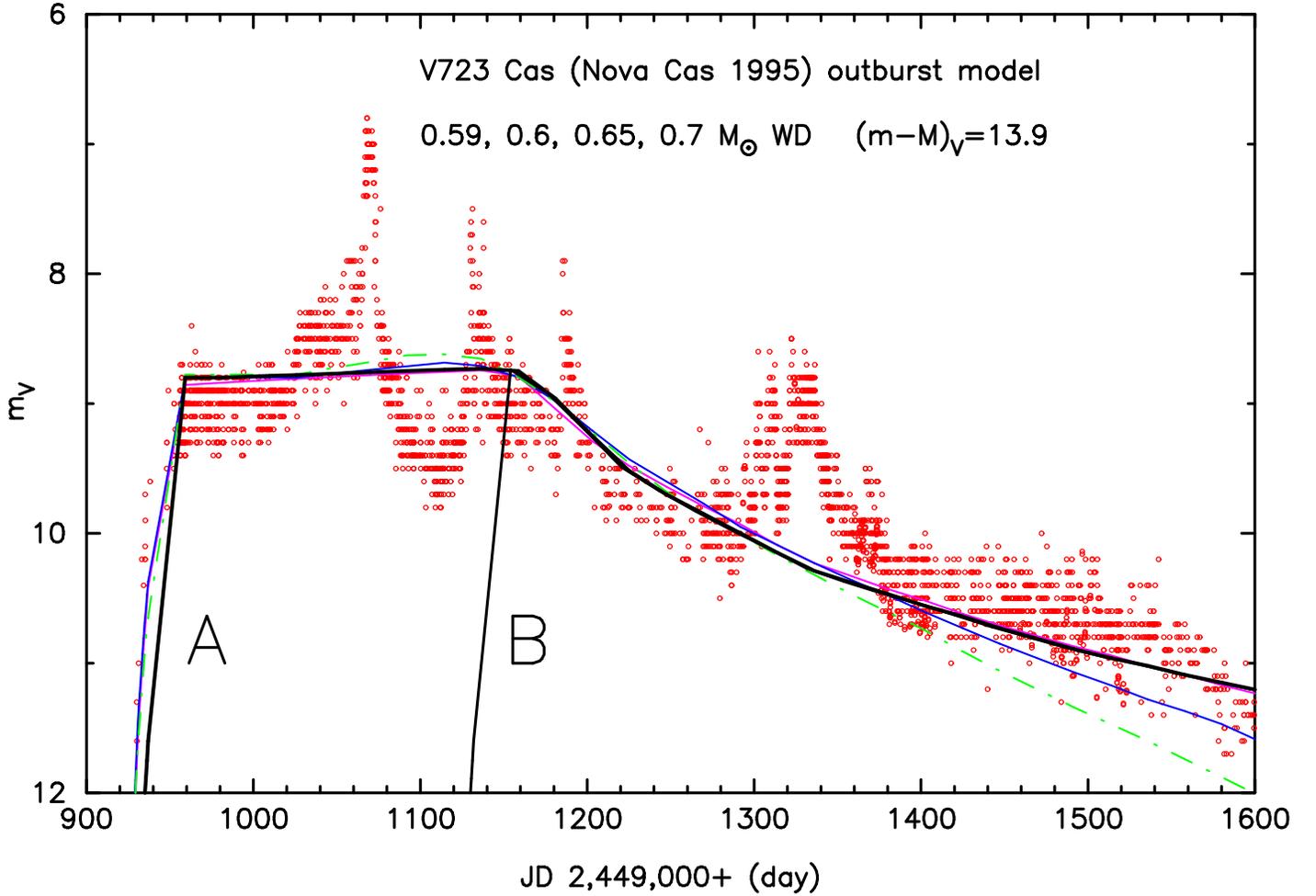}
\caption{
Calculated $V$ light curves plotted against time (JD 2,449,000$+$)
together with the observational points \citep[{\it small open circles},
taken from VSNET archive; see also][]{kiy04}.
A thick solid line denotes the $V$ light curve for a $0.59 ~M_\sun$
WD model, which is the best-fit model for V723 Cas.
Two thin solid lines represent $V$ light curves of 
$0.60$ and $0.65 ~M_\sun$ WDs.
A thin dash-dotted line denotes the $0.70 ~M_\sun$ WD.
The light curve of $0.60 ~M_\sun$ is almost overlapped with 
that of $0.59 ~M_\sun$.  The labels A and B correspond to
those in Figure \ref{hr_diagram_v723cas1995}.
\label{vmag_mmix_v723cas1995_early}}
\end{figure}

\begin{figure}
\plotone{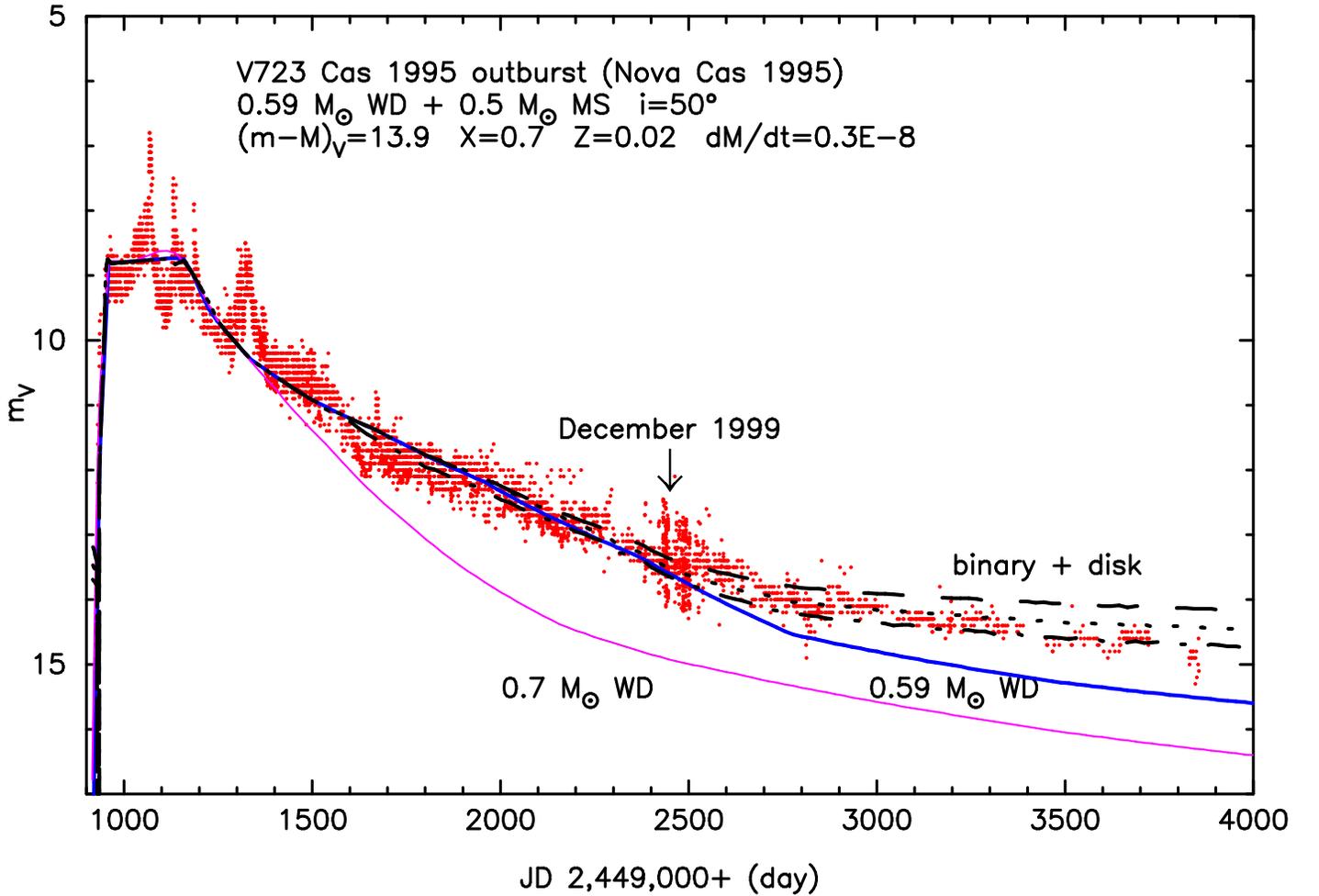}
\caption{
Same as in Fig. \ref{vmag_mmix_v723cas1995_early}, but for the 
full period.
A thick solid line denotes the calculated $V$ light curve 
of the $0.59 ~M_\sun$ WD photosphere only, while a thin solid line
denotes the $0.70 ~M_\sun$ WD photosphere.
Dotted, dash-dotted, and dashed lines are the calculated 
mean, minimum, and maximum orbital $V$ magnitudes of 
the $0.59 ~M_\sun$ WD photosphere, an accretion disk, 
and a companion star of $0.5 ~M_\sun$, respectively.
In the very late phase, the amplitude of the 
orbital modulations reaches about a magnitude, which is
consistent with the observational results by
\citet{cho03}. 
\label{vmag_mmix_v723cas1995}}
\end{figure}

\begin{figure}
\plotone{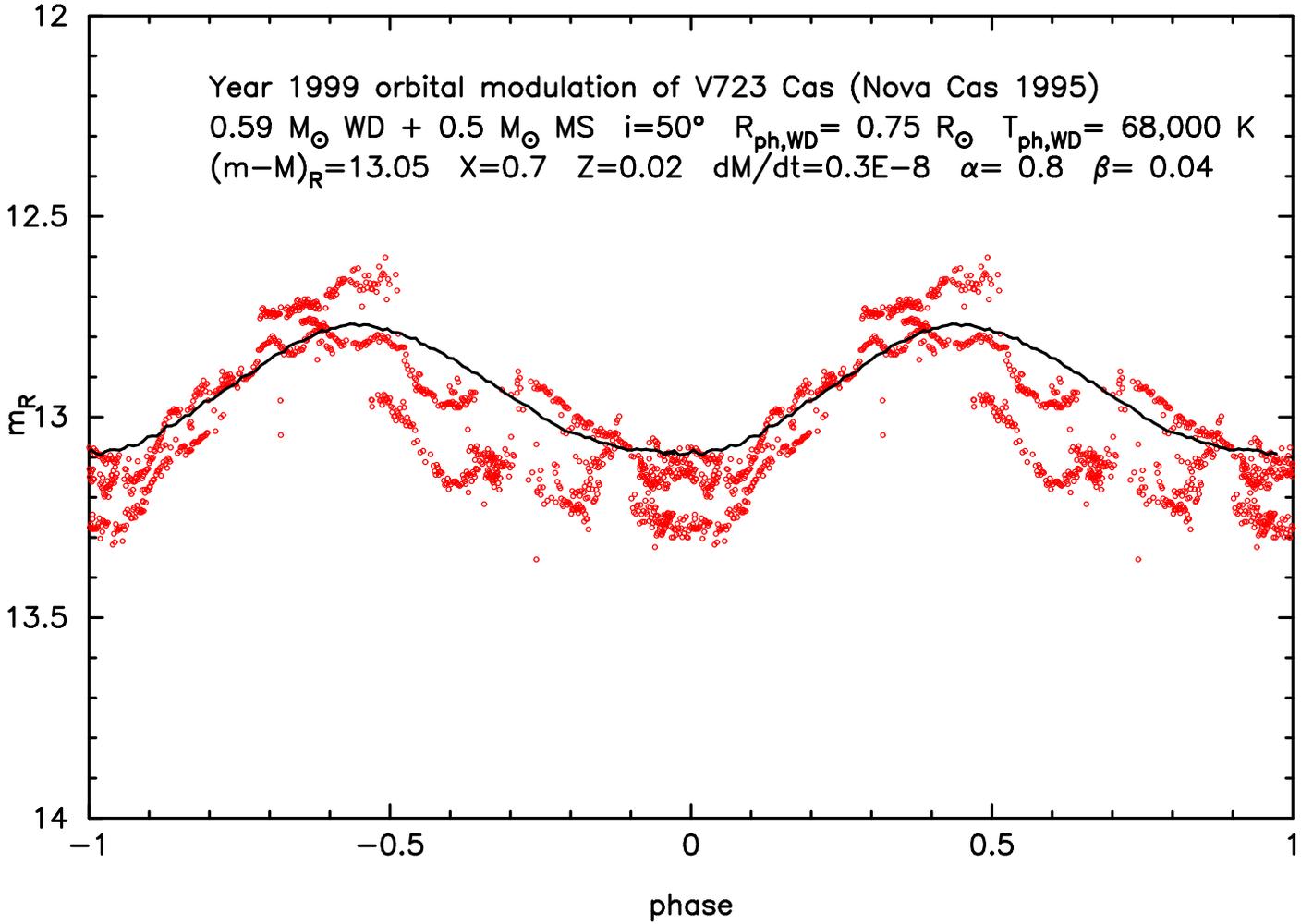}
\caption{
Calculated orbital Cousins $R_{\rm C}$ light curve of V723 Cas
plotted against the orbital phase.  
The orbital phase is repeated twice from $-1.0$ to $1.0$.
A solid line is the calculated $R_{\rm C}$ light curve while 
small open circles are Johnson $R$ observational points taken from
\citet{gor00} in 1999 September-December.
The inclination angle is
about $i= 50\arcdeg$ for our best-fit model.
\label{rmag_v723cas_late_orbital}}
\end{figure}

\begin{figure}
\plotone{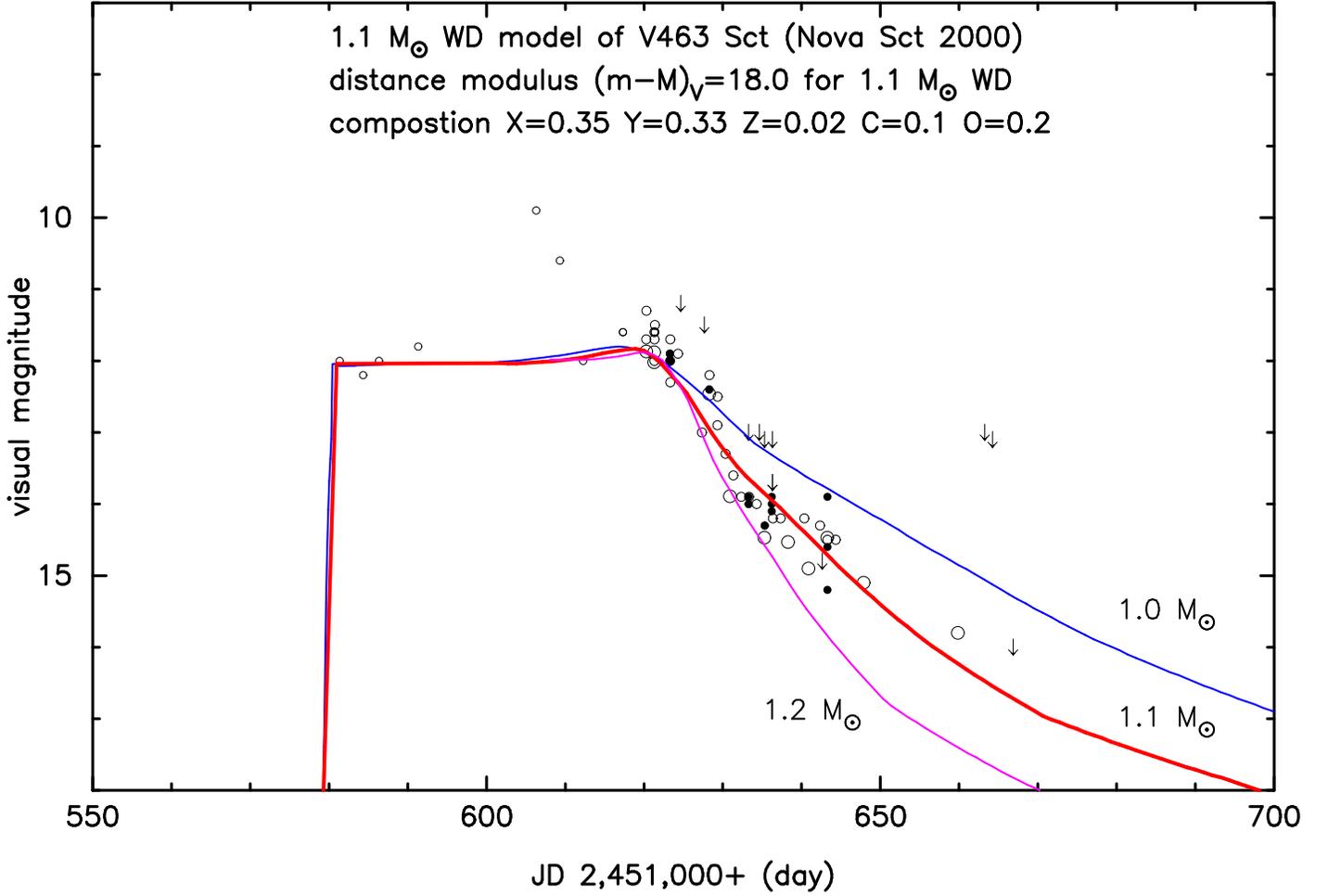}
\caption{
Same as in Fig. \ref{vmag_mmix_v723cas1995_early}, but for
the V463 Sct (Nova Sct 2000) outburst.
A thick solid line denotes the calculated $V$ magnitude
of the $1.1 ~M_\sun$ WD photosphere, the best fitted
curve with the observation.  Small open circles, middle-size
open circles, large open circles, and filled circles 
represent observational points of 
photographic, visual, $V$, and CCD magnitudes, respectively.
Arrows indicate an upper limit.
The apparent distance modulus is estimated to be
$(m-M)_V = 18.0$.
\label{vmag_outburst_v463sct}}
\end{figure}

     The premaximum halt works as a standard candle of the 
Eddington luminosity.  Strictly speaking, the Eddington
luminosity, $L_{\rm Edd} = 4 \pi c G M_{\rm WD} / \kappa$, is a
local variable because the opacity is a local variable.
Although the opacity varies largely, the photospheric luminosity,
$L_{\rm ph}$, is almost constant as shown in 
Figures \ref{hr_diagram_v723cas1995}.
Therefore, we regard that $L_{\rm ph}$ in the almost flat part
of the H-R diagram plays the role of the Eddington luminosity.
The corresponding flat peak (fp) absolute visual magnitude, 
$M_{V, {\rm fp}}$, 
depends on the WD mass, which is approximately given by
\begin{equation}
M_{V, {\rm ~fp}} \approx -1.53 ~(M_{\rm WD}/M_\sun) - 4.26,
\label{peak_luminosity_solar}
\end{equation}
for the solar composition of WD envelopes, or by
\begin{equation}
M_{V, {\rm ~fp}} \approx -1.75 ~(M_{\rm WD}/M_\sun) - 4.25,
\label{peak_luminosity_co}
\end{equation}
for carbon and oxygen enrichments of $C+O=0.30$,
where these two relations are valid for 
$0.6 \le (M_{\rm WD}/M_\sun) \le 1.3$.
These flat peak absolute magnitudes give a standard candle instead of
the Eddington luminosity.  Comparing the above flat peak 
visual magnitude $m_{\rm V, fp}$ with the maximum visual 
magnitude $m_{\rm V, max}$ (see Figs. 2 and 5), we may conclude 
that the nova outburst was 1.8 mag super Eddington for V723 Cas 
\citep[see also][]{cho98} and, at least, 2.0 mag
super Eddington for V463 Sct.  Our analysis can be extended 
to symbiotic novae.  For example, we have preliminary analyzed
the well-observed symbiotic nova PU Vul 
\citep[Nova Vul 1979; see e.g.,][]{cho97b} and
found that its very long (3000 days) flat peak is reproduced
with a less massive $0.55 ~M_\sun$ WD, the details of which 
will be published elsewhere.

     Our steady state wind models show the photospheric velocity 
$v_{\rm ph} \sim 70$~km~s$^{-1}$ of the $0.59 ~M_\sun$ WD 
and $v_{\rm ph} \sim 200$~km~s$^{-1}$ of the $1.1 ~M_\sun$ WD 
at the very early phase of the flat peak.
The wind velocity gradually increases up to 
$v_{\rm ph} \sim 500$~km~s$^{-1}$ ($0.59 ~M_\sun$ WD) and
$v_{\rm ph} \sim 1000$~km~s$^{-1}$ ($1.1 ~M_\sun$ WD) 
in the late phases of the nova outbursts.
This is roughly consistent with 
FWHM$=500$~km~s$^{-1}$ of the V723 Cas 1995 outburst 
\citep{rud02} and
FWHM$=990$~km~s$^{-1}$ of the V463 Sct 2000 outburst \citep{kat02t}.

     The ejected mass by winds is estimated to be
$\Delta M_{\rm wind} = \Delta M_{\rm ejecta} = 3.6 \times 10^{-5}
M_\sun$ for the $0.59 ~M_\sun$ WD.
This is roughly consistent with the observational estimate 
of $2.6 \times 10^{-5} M_\sun$ \citep{eva03}.
\citet{kat02t} suggest, from the presence of [\ion{O}{1}] lines,
that a large amount of mass has 
already been ejected before the optical maximum 
(2000 March 2) of V463 Sct.  This is also consistent with
our estimations of $\Delta M_{\rm wind} = 2.1 \times 10^{-5} M_\sun$
before the optical maximum and
$\Delta M_{\rm wind} = 5.7 \times 10^{-5} M_\sun$ before the
spectroscopic observation.

     Our main conclusions are summarized as follows:
\par\noindent
1. The visual magnitude of a nova attains its maximum 
and has a flat peak when the envelope mass is massive enough.
The duration of a flat peak
depends on the initial envelope mass at the ignition.
This flat peak can be interpreted as a premaximum halt.
\par\noindent
2. The WD mass can be determined by the decline rate 
in the decay phase if the composition of a nova envelope is known.  
\par\noindent
3. The premaximum halt
works as a standard candle of the Eddington luminosity.
Thus we are able to calibrate the absolute magnitude 
of nova light curves by fitting our calculated light curve.
\par\noindent
4. The degree of the super-Eddington luminosity reached by
a nova in the visual maximum is estimated by the premaximum halt:
1.8 mag for V723 Cas and 2.0 mag for V463 Sct.



\acknowledgments
     We thank T. Kato and S. Kiyota for providing us with 
V463 Sct and V723 Cas data in the VSNET archive, 
and also V. P. Goranskij for sending us the orbital light curve 
data of V723 Cas. We are also grateful to T. Iijima for stimulating
discussion on V723 Cas and to the anonymous referee for useful
comments. 
This research has been supported in part by the Grants-in-Aid for
Scientific Research (16540211, 16540219) 
of the Japan Society for the Promotion of Science.

\clearpage

\clearpage

\clearpage

\clearpage


\end{document}